\documentclass[twocolumn,preprintnumbers,
nofootinbib,prd]{revtex4}

\usepackage{graphicx}

\usepackage[hypertex]{hyperref}
\newcommand{\vev}[1]{\langle {#1} \rangle}

\newcommand{\ord}[1]{\mathcal{O}{(#1)}}

\newcommand{\gsim}{\gtrsim}
\newcommand{\beq}{\begin{equation}}
\newcommand{\eeq}{\end{equation}}

\begin{document}

\pagestyle{plain}

\preprint{MADPH-05-1430}

\preprint{UFIFT-HEP-05-28}

\title{Implications of Neutrino Mass Generation from QCD Confinement}

\author{Hooman Davoudiasl\footnote{\tt email: hooman@physics.wisc.edu}}

\affiliation{Department of Physics, University of Wisconsin,
Madison, WI 53706, USA}

\author{Lisa~L. Everett\footnote{\tt email: everett@phys.ufl.edu}}

\affiliation{Department of Physics, University of Florida,
Gainesville, FL 32611, USA}


\begin{abstract}

We consider the possibility that the quark condensate formed by
QCD confinement generates Majorana neutrino masses $m_\nu$ via
dimension seven operators. No degrees of freedom beyond
the Standard Model are necessary, below the electroweak scale. 
Obtaining experimentally
acceptable neutrino masses requires the new physics scale $\Lambda
\sim$~TeV, providing a new motivation for weak-scale discoveries
at the LHC. We implement this mechanism using a $Z_3$
symmetry which leads to a massless up quark above the QCD chiral condensate
scale. We use non-helicity-suppressed light meson rare decay data
to constrain $\Lambda$. Experimental constraints place a mild
hierarchy on the flavor structure of dimension seven operators and
the resulting neutrino mass matrix.

\end{abstract}
\maketitle


Non-zero neutrino masses $m_\nu$ provide the simplest and most
robust explanation of neutrino oscillation data from a multitude
of experiments.  However, generically, models of neutrino mass
require physics beyond the Standard Model (SM) either far above or
well-below the electroweak scale $m_W \sim 100$~GeV.  Most of
these models are based on the seesaw mechanism \cite{seesaw}, in
which Majorana mass terms of the order $\Lambda_S\sim 10^{14}$~GeV
for the right-handed neutrinos are present in addition to the
usual electroweak Dirac mass terms, such that the tiny values of
$m_\nu$ are obtained from the ratio $m_W^2/\Lambda_S$.

Recently, classes of models have been proposed in which neutrino masses
arise from higher-dimensional operators suppressed by lower
scales $\Lambda \ll \Lambda_S$ (e.g. $\Lambda \sim 10$~TeV)
\cite{Chacko:2003dt,Chacko:2004cz,Davoudiasl:2005ks}.  A key ingredient is
the inclusion of new physics in the infrared,
typically well-below 1~GeV.  The new IR sector gives rise to novel
astrophysical signatures that have been recently studied
\cite{Davoudiasl:2005fd,Goldberg:2005yw}.

Given the typical scales involved in this class of models, it is
interesting to consider using the natural SM electroweak scale
$\Lambda \sim$~TeV and the scale $\Lambda_\chi \sim 100$~MeV of
chiral symmetry breaking in Quantum Chromo-Dynamics (QCD) to
generate $m_\nu$ of ${\cal O}(0.1)$~eV. In what follows, we
explore this possibility and outline the requirements that yield a
consistent scenario.

We first note that this general framework has also been considered by
Ref.~\cite{Thomas:1992hf}, using a similar line of reasoning and based on
new global $U(1)$ symmetries.  Although we share the general features of the
scenarios put forward in Ref.~\cite{Thomas:1992hf}, we construct a simple
model that satisfies the necessary requirements via a new $Z_3$ discrete
symmetry.  We also consider additional constraints
on these models from meson decay data, not considered in
Ref.~\cite{Thomas:1992hf}, which lead to stronger bounds on
the scale of new physics $\Lambda$.
If the active neutrinos
have Majorana masses, no new degrees of
freedom beyond the minimal SM are required below the electroweak scale.

This scenario predicts that the mass of the up quark is zero above
the scale of chiral symmetry breaking in QCD.
We will later comment on the consistency of a massless up quark
in light of the recent lattice results \cite{Nelson:2003tb,Aubin:2004fs}
and its implications for the strong CP problem.

Let us begin by outlining our theoretical framework. In an effective theory
below $\Lambda \sim $~TeV, suppression of lepton number violation, and hence
Majorana neutrino masses, requires that we forbid the dimension-5
operator\footnote{We assume $\ord{1}$ coefficients for operators in our
discussion, unless otherwise specified.}
\beq \mathcal{O}_H\sim \frac{(H L)(H L)}{\Lambda}\,
\label{OH}
\eeq
where generation indices are suppressed.  Given this assumption,
a simple way to do this is to impose
a discrete $Z_3$ symmetry \cite{Davoudiasl:2005ks},
under which the lepton doublet $L$ has charge $+1$ and the Higgs
doublet $H$ is neutral.

At this point, we would like to make a comment regarding the
nature of the scale $\Lambda$.  In our work, we will only assume
that $\Lambda$ is a scale of new physics and not necessarily
a cutoff scale where quantum gravity effects appear.
Thus, we do not expect a breakdown of all
global symmetries at or above $\Lambda$.  However,
if $\Lambda$ is treated as a cutoff scale, then one must impose
additional symmetries to suppress baryon
number violation and other experimentally forbidden processes,
as well \cite{Davoudiasl:2005ks}.

The SM fermions get their mass from Yukawa interactions of the
form
\beq
\mathcal{O}_Y \sim H {\bar f_L} f_R,
\label{OY}
\eeq
which couple the left- and the right-handed fermions $f_L$ and $f_R$,
respectively.  We assign $Z_3$ charge $+1$ to all
right-handed charged lepton fields
$e^{i}_R\,$, $i=1,2,3$, in the SM, so that the
Yukawa interactions (\ref{OY}) are allowed for them.

Schematically, we are interested in generating
Majorana masses $m_\nu \sim 0.1$~eV
from the dimension-7 operator $\mathcal{O}_q$ of the form \cite{Dirac}
\beq
\mathcal{O}^M_q \sim \frac{[({\bar Q_L}\,q_R)\cdot L]\, (H
L)}{\Lambda^3}\,,
\label{Oq}
\eeq
in which $Q_L$ is a left-handed quark doublet and $q_R$ is a
right-handed up-type quark.  The combination ${\bar Q_L}\,q_R$ is
a $\bar{2}$ of SU(2)$_L$ and has U(1)$_Y$ hypercharge $+1/2$.  We
must arrange for $\mathcal{O}_q^M$ to be $Z_3$ neutral if it is to be
allowed in our theory.  A simple way to achieve this is to endow $q_R$
with $Z_3$ charge $+1$.  However, this will forbid writing down
the Yukawa term (\ref{OY}) for $q_R$ and hence this quark remains
massless at scale $\Lambda$. Since the lightest quark in the SM is
the up quark, we will hereafter assume
that $q_R = u_R$ and thus the
up quark remains massless at the cutoff
scale: $m_u(\Lambda) = 0$.

So far, we have succeeded in forbidding $\mathcal{O}_H$ in Eq.~(\ref{OH})
and allowing the operator
\beq
\mathcal{O}_u^M = y_{ijk}\frac{[({\bar
Q_L^i}\,u_R)\cdot L^j]\, (H L^k)}{\Lambda^3}\,,
\label{Ou}
\eeq
where $y_{ijk} \sim 1$.
As a result of QCD confinement, the light
quarks $(u, d)$ in the SM form a condensate that breaks the global
SU(2)$_L$$\times$SU(2)$_R$ chiral symmetry of strong interactions.
In particular, $\vev{{\bar u_L} u_R} \neq 0$, which implies that
below the electroweak scale, Eq.~(\ref{Ou}) contains a
term
\beq
\mathcal{O}_{\nu\nu}^M = y_{1jk}\frac{[\vev{{\bar
u_L}\,u_R}\, \vev{H}]}{\Lambda^3} \nu^j \nu^k\,.
\label{Onunu}
\eeq
We thus require
\beq
m_\nu \sim \frac{\vev{{\bar u_L}\,u_R}\,
\vev{H}}{\Lambda^3}\,.
\label{mnu}
\eeq
For $\vev{{\bar u_L}\,u_R}
\simeq (200~{\rm MeV})^3$, $\vev{H} \equiv
v/{\sqrt 2} \simeq 174$~GeV, as required by
the SM, and $m_\nu = {\sqrt{\Delta m_{\rm atm}^2}}
\simeq 0.06$~eV, we get $\Lambda \simeq 3$~TeV.
Since established data only allow small variations in the 3 input
parameters that set $\Lambda$, our prediction for the scale
of new physics is unambiguous.

We saw that our simple construct forbids the usual up quark mass term,
and hence this mechanism appears to require $m_u(\Lambda) = 0$.  A
massless up quark has long been invoked as a possible resolution
of the strong CP problem.  This is because the CP violating
angle $\theta$ in the QCD Lagrangian is only defined up to the
phase of the quark-mass-matrix determinant.  With a zero
eigenvalue, the phase becomes undefined and the $\theta$-angle can
be rotated away.  However, recent lattice QCD calculations
\cite{Nelson:2003tb,Aubin:2004fs}
seem to show that the up quark is
not massless, in apparent conflict with our construct.
Next, we will argue that this is
not necessarily the case.

What has been shown by lattice calculations is that setting $m_u =
0$ at a scale of order $\Lambda_{\rm QCD} \sim 100$~MeV cannot be
compensated by a contribution from the next to leading order
chiral Lagrangian.  This contribution was shown by Kaplan and
Manohar \cite{Kaplan:1986ru} to induce an effective up quark mass
and originates in a redundancy of the chiral Lagrangian
formulation \cite{Banks:1994yg}. Nonetheless, there is an
additive non-perturbative
contribution to $m_u$ \cite{Banks:1994yg}, due
to QCD instantons, that generate $m_u \neq 0$ at $\Lambda_{\rm
QCD}$, even if we set $m_u = 0$ at the cutoff scale $\Lambda \sim
1$~TeV. This contribution has a form similar to that of the
Kaplan-Manohar ambiguity, but is physically of a different origin
\cite{Banks:1994yg,Srednicki:2005wc}.  Therefore, we hold that
the requirement $m_u (\Lambda) = 0$ is not
necessarily in conflict with the lattice results.

Here, we would like to add that the operator in Eq.~(\ref{Ou})
contributes to the up-quark mass at 1-loop level.  The size of
this contribution can be estimated and is of order $\delta m_u
\sim \vev{H} m_\nu/\Lambda \sim 10^{-10}$~MeV.  In this estimate,
we have used $m_\nu \Lambda^2$ as the size of the neutrino loop.
The $\theta$-angle in QCD is constrained by the electric dipole
moment (EDM) of the neutron which is proportional to $m_u$
\cite{Baluni:1978rf,Crewther:1979pi}.  Since the contribution
$\delta m_u$ is suppressed by $10^{-10}$ compared to the usual
value of $m_u \sim 1$~MeV, the neutron EDM experimental bound is
consistent with $\theta \sim 1$ and hence the strong CP problem is
resolved for all practical purposes.  Note that 
the much larger instanton-generated mass of order 100~MeV, 
mentioned above, is a real contribution to the non-perturbative 
renormalization of the light quark masses and does not 
affect the resolution of the strong CP problem 
\cite{Srednicki:2005wc}.

We now examine the quantum stability of the
$Z_3$ symmetry we have employed in the above framework.
To this end, we inquire
whether this symmetry is anomaly-free in the SM.  The
fermions that are charged under $Z_3$ are $L^i$, $e^i_R$, and
$u_R$.  As both the number of leptonic generations and the number
of QCD colors are equal to 3, all triangle anomalies related to
SU(2)$_L$, U(1)$_Y$, and gravity are zero {\it mod} 3.  The only anomaly
is from the triangle with two SU(3)$_c$ gluons
and one $Z_3$ vertex, since only $u_R$ has $Z_3$ charge.

We see that the $Z_3$ symmetry is not exact at the quantum level
and cannot be imposed as a {\it gauge} symmetry at scale
$\Lambda$.  In principle, the anomaly can be canceled by
introducing new fermions near the cutoff scale.  This possibility
will lead to the presence of new massless SU(3)$_c$ charged
fermions in the theory.  These fermions stay massless as long as
the $Z_3$ we have imposed is not broken and can only have masses
of order $m_\nu$ from chiral symmetry breaking of QCD. This is in
stark conflict with nearly all experimental data. Thus, the only
way to have these fermions in our theory is to push their masses
above $\sim 100$~GeV, where they decouple from present data.
However, this would require $Z_3$ to be spontaneously broken at
$\sim 100$~GeV, which means that $m_\nu$ would no longer be
protected down to $\Lambda_{\rm QCD}$, negating the purpose of
having this symmetry in the first place.

In our treatment, we will not attempt to resolve the issue of
anomaly cancellation. In 4-d, 
the anomaly of our $Z_3$ symmetry 
suggests that it must be thought of as an accidental
symmetry.  That is, like baryon number symmetry in the SM, it
arises from gauge invariance and renormalizability of the
underlying theory.  Thus, we expect that a new gauge symmetry
beyond the SM to exist above scale $\Lambda$ in a UV completion of
our framework.  

Another interesting possibility could be provided 
by anomaly cancellation in extra-dimensional models.  
One could entertain a scenario in which the $Z_3$ anomaly of the 
``visible sector'' is canceled by contributions from 
other fields that are localized on various  
defects in extra dimensions.  For example, a 5-d theory with three 
identical 3-branes containing the same field 
content as our framework will 
have no $Z_3$ anomaly.    
Depending on the details of compactification and the 
underlying geometry of the extra 
dimensions, this scenario could result 
in the appearance of collider and 
other high energy experimental 
signatures.  We will not attempt to construct 
a UV complete theory that induces
our $Z_3$ at low energies; 
this is beyond the scope of this work which
focuses on the phenomenological implications of neutrino mass
generation from QCD confinement.

The central features of the mechanism studied here are encoded in
Eqs.~(\ref{Ou}) and (\ref{Onunu}): the QCD chiral-condensate and
$\vev{H}$, all SM ingredients, can be incorporated into an
effective suppressed mass term which yields acceptable neutrino
masses. We emphasize that because this operator is exclusively
constructed out of SM fields, {\it the size of the scale where new
physics emerges is not arbitrary and must be at a few TeV.}
Therefore, the above mechanism for neutrino mass generation
motivates new physics near the electroweak scale, {\it
independently} of the gauge hierarchy problem. Examples of
specific scenarios with new scalars at TeV energies have been
presented in Ref.~\cite{Thomas:1992hf}, using a $U(1)$ global
symmetry. Consequently, we expect such new physics, relating
neutrino masses to QCD, will be accessible at the LHC.

Here, we turn to the question of experimental constraints on this
mechanism.  The operator $\mathcal{O}_u^M$ in (\ref{Ou}) can also
lead to new decay channels for charged pseudoscalar mesons, with
$Q_L^i = d_L^i$:
\beq \mathcal{O}_{du} =
\left(\frac{\vev{H}}{\Lambda}\right)\frac{({\bar d_L^i}\,u_R) \,
e^j\, \nu^k}{\Lambda^2}\,,
\label{Odu}
\eeq
where we have set
$y_{ijk}=1$ for simplicity and consider bounds on the effective
value of $\Lambda$, taking all the relevant coefficients to be
unity. Ref.~\cite{Thomas:1992hf} considered this possibility for
the lepton number violating decay $\pi^+ \to \mu^+ {\bar \nu}$,
for which the helicity of the $\mu^+$ has the wrong sign compared
to the SM decay $\pi^+ \to \mu^+ \nu$.  The total muon
polarization depends on the scale $\Lambda$ as: $1 - |P_\mu|
\propto 1/\Lambda^6$. It was found that $\Lambda \gsim 1$~TeV, given
the available data~\cite{Thomas:1992hf}: $|P_\mu|>0.9959$.
One can easily verify that using the most recent 90\%
C.L. bound from PDG, $|P_\mu|>0.9968$, would only provide a tiny
improvement, resulting in effectively the same bound as before.

However, we can achieve much stronger bounds on the
the cutoff scale if we consider the contribution of the $\mathcal{O}_{du}$
to the partial decay width of a light pseudoscalar
$P_i \to e^+ {\bar \nu}_x$,
where $P^+_i = \pi^+, K^+, B^+$, for $i=1,2,3$, respectively;
$\nu_x$ is an active neutrino.  Note that the decay channels
including $e^\pm$ are severely helicity suppressed in the SM.
However, lepton number violating decays mediated by $\mathcal{O}_{du}$
are not helicity suppressed. The current Particle Data Group
(PDG)~\cite{Eidelman:2004wy} bounds are
on the SM processes $P^+_i \to e^+ \nu_e$.  However, since the
quantum numbers of the final state neutrino is not measured, 
these bounds constrain $P^+_i \to e^+ \nu_x$, where $\nu_x$ is 
any type of neutrino.  Thus, the
helicity unsuppressed processes mediated by $\mathcal{O}_{du}$ will
contribute to the measured branching fractions 
Br($P^+_i \to e^+ \nu_e$).

The partial width $\Gamma_i$ for $P_i \to e^+ {\bar \nu}_x$
mediated by $\mathcal{O}_{du}$ is given by
\beq
\Gamma_i = 3\left(\frac{v^2}{128\, \pi \Lambda^6}\right)
f_i^2 \mu_i^2 m_i.
\label{Gam}
\eeq
In the above, $m_i$ is the mass of $P^+_i$, the factor 3
refers to the contribution of 3 diagrams corresponding
to the unobserved neutrino flavors, and $v=246$~GeV ($\vev{H}\equiv
v/\sqrt{2}$).  We also have
$\mu_1 = m_\pi^2/(2 {\bar m}),
$
with ${\bar m} = (m_u + m_d)/2 \simeq 5$~MeV; $m_u$
and $m_d$ are the up and down quark masses, respectively.
Hence $\mu_1 = 14 m_\pi$.  For the $i=2$,
$\mu_2 = m_K^2/(m_s + {\bar m})$, with $m_s \simeq 100$~MeV
the strange quark mass; $\mu_2 \simeq 5 m_K$.  We also get
$\mu_3 \simeq m_B$, for the $B^+$ meson, where the
$b$ quark mass $m_b \simeq m_B$.  The $f_i$ are the pseudoscalar meson
decay constants: $f_1=130$~MeV, $f_2=160$~MeV, and
$f_3=180$~MeV.

From PDG~\cite{Eidelman:2004wy}, we have
\beq
{\rm Br}(\pi^+ \to e^+ \nu_e) = (1.230 \pm 0.004)\times 10^{-4},
\label{pi}
\eeq
\beq
{\rm Br}(K^+ \to e^+ \nu_e) = (1.55 \pm 0.07)\times 10^{-5},
\label{kaon}
\eeq
and
\beq
{\rm Br}(B^+ \to e^+ \nu_e) < 1.5\times 10^{-5} \;
(90\% \;{\rm C.L.}).
\label{bmeson}
\eeq
We require that the contribution to the above branching fractions
from Eq.~(\ref{Gam}) is smaller than the $1-\sigma$ uncertainty
on the measured branching fraction, or smaller than the bound, in the
case of $B^+$.  The following bounds
\beq
\Lambda > 8.6, 9.8, 2.7 \; {\rm TeV}
\label{bounds}
\eeq
are obtained for the $\pi^+$, $K^+$, and $B^+$, respectively.
Note that these bounds are weaker than those obtained from usual
(pseudo)scalar leptoquarks, which are typically of ${\cal
O}(100\,{\rm TeV})$ for ${\cal O}(1)$ coefficients
\cite{Altarelli:1997qu,Shanker:1982nd}. The bounds derived here
are weaker because these operators do not interfere with SM
processes, which is not the case for the operators considered in
\cite{Altarelli:1997qu,Shanker:1982nd}.

The strongest
bound arises from K mesons; $\Lambda \simeq 10$~TeV.
This constraint leads to neutrino masses which are too small to
account for the observed atmospheric neutrino mass-squared
difference. However, this bound was obtained by assuming all
coefficients $y_{ijk}$ of ${\cal O}(1)$. In practice, such
coefficients should have an intrinsic flavor dependence.  The
meson decays $P_i\rightarrow e^+\bar{\nu}_x$ constrain the
coefficients $y_{i1k}$, with $i=1$ for the pions, $i=2$ for the
kaons, and $i=3$ for the $B$ mesons, while the neutrino mass
matrix involves the coefficients $y_{1jk}$ (see Eq.~(\ref{Ou})).
Therefore, the bound from the $K$ mesons can be alleviated if
there is a flavor-dependent hierarchy between the $i=1$ and $i=2$
operators of ${\cal O}\sim (3\,{\rm TeV}/10\,{\rm TeV})^3\approx
(1/30)$, where 3 TeV is the mass scale required for realistic
neutrino masses as derived above.  However, we still have the
bound from the pions, which is similar and cannot be addressed by
flavor hierarchy amongst the operators.  To see this, note that
the operator which contributes to $\pi^+$ decay also contributes
to the neutrino mass matrix up to an $SU(2)_L$ transformation, so
gauge invariance does not allow further tuning of coefficients.

To have a consistent framework, we thus require that the
coefficient $y_{11k}$ in Eq.~(\ref{Ou}) be suppressed at the level
$(3/9)^3 \approx 1/30$, similar to the suppression required from
the $K$ decay data.  This, given the above discussion, also leads
to suppression of neutrino mass matrix elements $M_{1k}$.
Therefore, the consistency of our framework seems to suggest
hierarchical neutrino masses.  Hence, we see that the meson decay
data provide some guidance for constructing a UV completion of
our effective theory at scale $\Lambda$.  Since the scale of new
physics is in the TeV regime, this framework could provide an
interesting avenue of exploration for weak scale model building.

In conclusion, we studied the consequences of generating neutrino
masses from QCD confinement via higher-dimensional operators.
Neutrino masses are suppressed by the ratio of the QCD chiral
condensate to the scale of new physics. The higher dimension
operators consist only of SM fields. Consequently, the scale of
new physics cannot be larger than a few TeV, in order to generate
acceptable neutrino masses. This strongly suggests that the LHC
will probe the new physics that relates non-perturbative QCD
dynamics and neutrino masses.  We considered a model that requires
a massless up quark above the QCD confinement scale. We argued
that this is not necessarily in conflict with lattice results and
may resolve the strong CP problem. Below the electroweak scale,
the model reduces to the field content of the minimal SM.  We used
rare decays of light pseudo-scalar mesons to place bounds on the
effective scale of new physics.  The experimental constraints
suggest a quark and lepton flavor-dependent ${\cal O}(10^{-1})$
hierarchy among the coefficients of the higher dimension
operators.  Since the neutrino mass matrix is generated from these
higher dimensional operators, a generic consistency condition for
our framework is a hierarchical pattern of neutrino masses and
mixings.

\acknowledgments

We thank V. Barger, A. de Gouv\^{e}a,
T. Han, P. Huber, R. Kitano,  H. Logan, K. Melnikov,
M. Peskin,  and E. Witten for discussions. 
We are especially grateful to D. Chung, P. Langacker, and C. Thorn
for many useful comments.  
This work was supported in part by the United States
Department of Energy under Grant Contracts No. DE-FG02-95ER40896
(Wisconsin) and No. DE-FG02-97ER41209 (Florida).   H.D. was also
supported in part by the P.A.M. Dirac Fellowship, awarded by the
Department of Physics at the University of Wisconsin-Madison. L.E.
is also supported by a L'Oreal for Women in Science 2005
Postdoctoral Fellowship.


\end{document}